 \documentclass[twocolumn,aps]{revtex4}   
\usepackage{graphicx}

\newcommand{\beq}{\begin{equation}}
\newcommand{\enq}{\end{equation}}
\newcommand{\bqa}{\begin{eqnarray}}
\newcommand{\eqa}{\end{eqnarray}}
\newcommand{\bqas}{\begin{eqnarray*}}
\newcommand{\eqas}{\end{eqnarray*}}
\newcommand{\bea}{\begin{array}}
\newcommand{\ena}{\end{array}}

\begin{document}
\draft
\widetext
 
\title{Quantum entanglement and the self-trapping  transition in polaronic systems}

\author{Yang Zhao$^1$, Paolo Zanardi$^2$, and Guanhua Chen$^1$}
\address{
$^1$Department of Chemistry, University of Hong Kong,
Pokfulam Road, Hong Kong, P.R.~China\\
$^2$Institute for Scientific Interchange, Villa Gualino, Viale 
Settimio Severo 65, I-10133 Torino, Italy
}
\date{\today}
\widetext
\begin{abstract}
We revisit from a quantum-information perspective a classic problem of polaron theory
in one dimension. In the context of the Holstein model we show that a simple analysis of quantum entanglement between
excitonic and phononic degrees of freedom allows one to effectively characterize both the small and large polaron regimes
as well as the crossover in between. The small (large) polaron regime 
corresponds to a high (low) degree of bipartite quantum entanglement between 
the exciton and the phonon cloud that clothes the exciton. 
Moreover, the self-trapping transition is 
clearly displayed by a sharp drop of
exciton-phonon entanglement.

\end{abstract}

\maketitle
\narrowtext

\section{Introduction} 

Almost three quarters of a century have passed since the concept of polaronic 
self-trapping was first conceived by Landau \cite{lan,pek}, and yet 
many questions remain unanswered regarding 
some of the most simplistic polaron Hamiltonians. 
In this paper  we will approach this classic problem from the 
contemporary point of view of quantum 
entanglement \cite{ent}. The latter concept plays a central role 
in the burgeoning field of quantum information
in that quantum entanglement represents the key physical resource 
at the basis of quantum information protocols
\cite{qip}. More recently, a growing amount 
of attention has been devoted to analyses of quantum entanglement
in many-body systems undergoing quantum phase transitions \cite{qpt-ent}.

An emerging picture is that various 
entanglement measures, such as the two-qubit concurrence and the
entanglement entropy of a subsystem, provide different, 
often complementary physical insights 
onto the nature of
many-body wave functions. Tools drawn from quantum information theory 
provide a deeper, previously unavailable 
understanding of quantum correlations 
and their qualitative changes across boundaries of different 
phase-diagram regimes in these many-body systems. 
Moreover, inspired by ideas from quantum-information theory, 
novel and efficient computational schemes
for studying quantum many-body systems have been devised \cite{comp-scheme}. 

The study to be carried out in this paper aims to exemplify
the ongoing efforts bridging
the new field of quantum information theory 
and established areas of condensed matter physics.
We will show that an analysis of quantum entanglement between electronic (or excitonic) and phononic 
degrees of freedom in a polaronic model allows one to characterize the polaron phase diagram in a strikingly simple fashion.
In particular, the self-trapping transition, i.e.,  a crossover between 
the so-called small and large polaron regimes, is readily captured 
by the behavior of the linear entropy of the excitonic reduced density matrix. 
We will show that the small (large) polaron regime
corresponds to a high (low) degree of quantum entanglement between the exciton and its phononic environment. Moreover,
the self-trapping transition is clearly displayed 
by a sharp drop of exciton-phonon entanglement.

It is worthwhile to stress that, 
at variance with most of the other studies
of this type,  
the particular form of 
quantum entanglement we are going to analyze 
is
between systems of distinct
physical {\em nature}: the exciton, a finite-dimensional system, and the
phonons, a bosonic bath. 
To emphasize this fact we will refer to 
the bipartite quantum entanglement between the exciton and its phonon
environment in the polaron problem
as {\em hetero}-entanglement. This situation bears resemblance to those
in {\em decoherence} studies where a system under examination, 
e.g., a qubit, is coupled with its environmental degrees of freedom, which spoils the purity of the system state.
From this point of view, the polaronic entanglement to be 
analyzed in this paper can be viewed as a measure of the  
decoherence of the excitonic (phononic) state induced by the coupling with the lattice phonons (excitons).

This paper is structured as follows. In Section II, 
the Holstein model is introduced, and 
the corresponding phase diagram of polaronic self-trapping is presented.
In Section III, we evaluate hetero-entanglement between the exciton
and the phonon bath, and examine its relation to polaronic self-trapping.
Discussions are given in Section IV, in which we show that 
excitonic superradiance, 
viewed as a form of intra-exciton entanglement between 
spatially distinct excitonic modes,
complement
the exciton-phonon entanglement. 


\section{Self-Trapping and Phase Diagram of the Holstein Model}

We first introduce a Frenkel-exciton model
Hamiltonian, also known as the Holstein molecular crystal model, which 
describes a lattice of
two-level molecules interacting with a bath consisting of nuclear
(intramolecular, intermolecular, and solvent) degrees of freedom:     
\begin{eqnarray}
\hat H = \sum_n \Omega_n({\bf q}) B_n^{\dagger} B_n + \sum_{m,n}^{m \neq n}
J_{mn}({\bf q}) B_m^{\dagger} B_n + \hat H^{ph} \quad .
\label{eqham1}
\end{eqnarray}
Here $B_n$ ($B_n^{\dagger}$) are exciton annihilation (creation)
operators for the $n$th molecule,
$\hat H^{ph}$ is the bath (phonon)
Hamiltonian,
and ${\bf q}$ represents the complete set of nuclear coordinates.
The excitonic operators satisfy the hard-core bosons relations
$(B_n^\dagger)^2=B_n^2=0,\,[B_n, B_m^\dagger]=\delta_{n m}.$
It follows that for each site label $n$ is associated to a two-level
system i.e., a {\em qubit}. Notice that in terms of the Pauli
matrices one has $B^\dagger=\sigma^+,$ and $B=\sigma^-.$
Despite this latter notation is the standard one in quantum information, 
in order to keep in line with the vast 
polaronic literature,  
we will stick to the $B$ operators. Interested readers 
should not have any problem 
in translating the formulas in the Pauli matrices notation.

Exciton-phonon interactions originate from 
dependence
of molecular frequencies $\Omega_n$
and the intermolecular couplings $J_{mn}$
on nuclear coordinates ${\bf q}$.  
We adopt the Hamiltonian Eq.~(\ref{eqham1})
with the Einstein phonon
Hamiltonian
$\hat{H}^{ph} = \sum_n \hbar \omega_0 b_n^\dagger b_n,
$
where  $b_n^\dagger$ creates
a phonon of frequency $\omega_0$ on site $n$, and we have one Einstein
oscillator per molecule.
Exciton-phonon interactions
enter through
the nuclear coordinate influence on both molecular
frequencies (diagonal coupling) and intermolecular interactions
(off-diagonal coupling).
Expanding $\Omega_n ({\bf q})$
to first order in
phonon coordinate $\bf {q}$, the first term of Eq.~(\ref{eqham1}) reads
$\sum_n \Omega_n ({\bf q}) B_n^{\dagger} B_n = \sum_n \Omega_n ({\bf q} =0)
B_n^{\dagger} B_n + \hat{H}^{diag} 
$
with the diagonal exciton-phonon coupling term
\begin{equation}
\hat{H}^{diag} = g \hbar \omega_0 \sum_n B_n^{\dagger} B_n ( b_n^{\dagger} +
b_n ),
\label{site}
\end{equation}
and $g$ is a dimensionless diagonal coupling constant.
Expanding $J_{mn} ({\bf q})$ to first order in phonon
coordinates, we write the second term of Eq.~(\ref{eqham1}) as, for example,
$\sum_{m \neq n} J_{mn} ({\bf q}) B_m^\dagger B_n =
\sum_{m \neq n} J_{mn} ({\bf q}=0 ) B_m^\dagger B_n + \hat{H}^{o.d.}$
with the transfer integral
$ J_{mn} ({\bf q}=0 ) = -J \delta_{n,m\pm 1}$ and the off-diagonal
coupling term \cite{munn,zhao94}
\begin{eqnarray}
\hat{H}^{o.d.}& =& \frac 1 2 \phi \hbar \omega_0
\sum_{nl} [ B_n^{\dagger} B_{n+1} ( b_l^{\dagger} + b_l)
(\delta_{n+1,l}-\delta_{nl})
\nonumber \\
& & +B_n^{\dagger} B_{n-1}
( b_l^{\dagger} + b_l) (\delta_{nl}-\delta_{n-1,l})].
\label{off}
\end{eqnarray}
The second term of
(\ref{off}) is the Hermitian conjugate of the first, and
we have assumed 
nearest-neighbor coupling 
of the antisymmetric type
with $\phi$ a
dimensionless parameter controlling
the off-diagonal coupling strength. Off-diagonal coupling
may adopt various forms \cite{zhao94,nago} other than the
antisymmetric type (\ref{off}), and can play important roles in electronic
properties of solid. In the theory of high-temperature superconductivity,
for example, it has been recently proposed \cite{nago}, that
off diagonal coupling modulates the hopping integral of the 
Zhang-Rice singlet and the superexchange interaction, and is 
especially relevant in the low-doping regime. 
Eq.~(\ref{site}) and Eq.~(\ref{off}), together with $\hat{H}^{ph}$      
and the zeroth-order intermolecular coupling term,
result in the generalized Holstein Hamiltonian $\hat{H}^{GH}$
(The original Holstein Hamiltonian contains diagonal coupling only)
\cite{holstein}:
\begin{eqnarray}
\hat{H}^{GH} & =& \sum_n \Omega_n ({\bf q} =0) B_n^{\dagger} B_n
+\hat{H}^{diag} \nonumber \\
&+&\sum_{mn}^{m \ne n} J_{mn} ({\bf q} =0) B_m^{\dagger} B_n +
\hat{H}^{o.d.}
+\hat{H}^{ph} .
\end{eqnarray}
 
There are two competing energy scales in the Holstein Hamiltonian in the absence
of off-diagonal exciton-phonon coupling, namely,
the lattice relaxation energy $g^2 \omega_0$ and the bare exciton bandwidth
$4J$.
Their ratio will be denoted the coupling strength
$\kappa=g^2 \omega_0/4 J$
which determines the size of the polaron as well as exciton-phonon correlations.
In typical molecular crystals,
$g^2~ \raisebox{-0.6ex}{$\stackrel{ <}{\scriptstyle \sim}$} ~1$,
in ionic crystals $g^2$ is large compared to unity,
and
in semiconductors $g^2$ is between the former two.
In anthracence, for example, $\kappa$
is about $0.4$, and in pyrene, about $0.8 \sim 1.6$ \cite{Matsui}.   
For strong exciton-phonon coupling ($\kappa \gg 1$),
solutions of the Holstein Hamiltonian
are known as small polarons
because the exciton-induced lattice distortion is
confined to essentially a single exciton site \cite{emin73}.
For weak exciton-phonon
coupling ($\kappa \ll 1$),
the spatial extent of the lattice distortion is significantly increased
and the resulting phonon-dressed exciton is called a large polaron.
The crossover from a large polaron to a small polaron
with increasing exciton-phonon coupling
(essentially often called the self-trapping transition) is
rather abrupt for large intermolecular coupling $J$.
In the limit of slow lattice motions,
adiabatic polaron theories admit
approximate solutions in the form of solitons.

The one-dimensional Holstein Hamiltonian with diagonal and off-diagonal
coupling
to Einstein phonons has been previously 
\cite{toy,thesis,zhaos,pekar}
modeled by
a variational wave function pioneered by Toyozawa (labeled as the Toyozawa ansatz
in Ref.~\cite{zhaos}):
\begin{equation}
|K \rangle = N^{-1} \sum_{n} e^{iKn} | \Lambda_n^K \rangle
\sum_{m}
\psi_{m-n}^K B_m^\dagger |0\rangle_{\rm e} \quad .
\label{113}
\end{equation}
Here $| K \rangle$ is the lowest energy polaron state with momentum $K$,
$ |0\rangle_{\rm e}$ is the exciton vacuum state, and $| \Lambda_n^K \rangle$
are phonon wave functions centered at site          
$n$ containing a coherent state on each site $n_2$ with
a displacement $\lambda_{n_2-n}^K$:
\begin{equation}
|\Lambda_n^K\rangle = \exp [ - \sum_{ n_2} (\lambda_{n_2 -n}^K
b_{n_2}^{\dagger} -\lambda_{n_2 -n}^{ K \ast} b_{n_2 } )] |0\rangle_{\rm ph}.
\label{coh}
\end{equation}        
$|0\rangle_{\rm ph}$ is the phonon vacuum state, and $| \Lambda_n^K \rangle$
is different from $| \Lambda_{n^\prime}^K \rangle$ only by a shift
of $n-n^\prime$ lattice constants. The parameters $\lambda_l^K$ and
$\psi_{l}^K$ are obtained variationally.
The phonon wave functions
$| \Lambda_n^K \rangle$ represent a lattice distortion forming a
potential well centered at $n$ and trapping the exciton with an amplitude
distribution of $\psi_{l}^K$.    
The Toyozawa Ansatz state (\ref{113}) is not normalized:
$\langle   K  |  K  \rangle  =  
\sum_{nm} e^{-i K n} \psi_m^K \psi_{m-n}^{K\ast} S_{n}^K $
where $ S_{n}^K $ is the Debye-Waller factor:  
$S_{n}^K \equiv  \langle\Lambda_{m}^K |\Lambda_{m-n}^K \rangle = \exp [N^{-1} \sum_q |\lambda_q^K|^2 (e^{iqn} -1)]
$
and
$\lambda_q^K$, are the Fourier transform of $\lambda_n^K.$
The variational methods are shown to be rather
efficient while remaining quantitatively accurate compared
with calculations involving far more expensive computational
resources \cite{jeck,trug}.

We introduce the phonon-traced exciton density matrix $\rho_{\rm e}^K$
for the state $|K\rangle$:
\begin{eqnarray}
\rho_{\rm e}^K :=
\langle K | K \rangle^{-1} 
{\rm Tr}_{\rm ph} (|K\rangle  \langle K| )
\label{114}
\end{eqnarray}
where ${\rm Tr}_{\rm ph}$ stands for tracing over the phonon degrees of 
freedom.
To calculate $\rho_{\rm e}^K$
we assume that
exciton-phonon coupling leads to the
formation of bands of collective exciton-phonon states, and
the many-body polaron wave function for the ground state band is given by
the Toyozawa Ansatz \cite{toy,thesis,zhaos}:
\begin{eqnarray}
(\rho^K_{\rm e})_{mm^\prime} = N^{-1} \langle K | K \rangle^{-1}
\sum_{n n^\prime} e^{iK(n^\prime-n)} \psi_{m-n}^K
\psi_{m^\prime-n^\prime}^{K \ast}
S_{n^\prime- n}^K .  \nonumber
\end{eqnarray}

\begin{figure}
\begin{center}
\rotatebox{0}{
\includegraphics[scale=0.54]{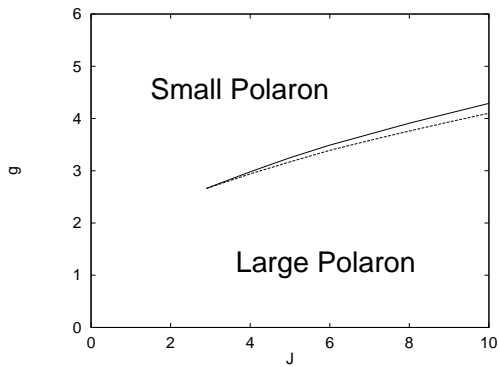}
        }
\end{center}
\caption{
The J-g phase diagram for the Toyozawa Ansatz. 
A thin sword-shaped regime identifies with the polaronic self-trapping
line, below which the effective mass of the polaron decreases drastically as
g is decreased. 
The top (solid) line of the sword-shaped regime indicates the
onset of bifurcation at $K = 0$ when the polaronic structure
goes through a sudden change as one travels
vertically downward in the phase diagram, and the bottom (dashed) line indicated where the state of the 
highest $K$ (in the vicinity of $K=0$) 
for which a discontinuous change in the polaronic 
structure is observed acquires two solutions as one travels
vertically upward.                  
} \label{phs}
\end{figure}            
 
In this paper we confine our attention to the case of diagonal exciton-phonon
coupling only. 
Localization of $(\rho^{K=0}_e)_{mm^\prime}$
is determined by the combined effect
of electronic confinement $\psi_{n}^{K=0}$ and 
the overlap of the adjacent nuclear wave functions $S_n^{K=0}$.
An efficient iterative procedure is employed to identify optimized
$\lambda_l^K$ and
$\psi_{l}^K$ \cite{thesis}.                    
Computation typically starts from the J=0 axis where exact solutions
exist. A convergent solution at one point in the phase diagram
is used to initiate a variation at a neighboring point. Reversibility
of the calculation and uniqueness of the solution are checked 
via altering initializations of the iterative procedure
as the parameter space is mapped. 
A phase diagram for the Toyozawa Ansatz 
spanned by $J$ and $g$ is presented in Figure 1 which displays a thin
sword-shaped region identified with the polaronic self-trapping. 
The top (solid) line of the sword-shaped regime indicates the 
onset of bifurcation at $K = 0$ when the polaronic structure
goes through a sudden change as one travels
vertically downward in the phase diagram, and the bottom (dashed) line 
indicated where the state of the highest
$K$ (in the vicinity of $K=0$),  
for which a discontinuous change in the polaron structure 
is observed, acquires two solutions as one travels 
vertically upward. 
Outside the sword-shaped area, solutions to self-consistent
variational equations are unique and independent of how the iterative procedure
is initialized. The polaronic structure for 
(J, g) points above (below) the sword-shaped area is traditionally identified as
small (large) polarons.

\section{Hetero-Entanglement and Polaronic Self-Trapping}

\begin{figure} 
\begin{center}
\rotatebox{0}{ 
\includegraphics[scale=0.67]{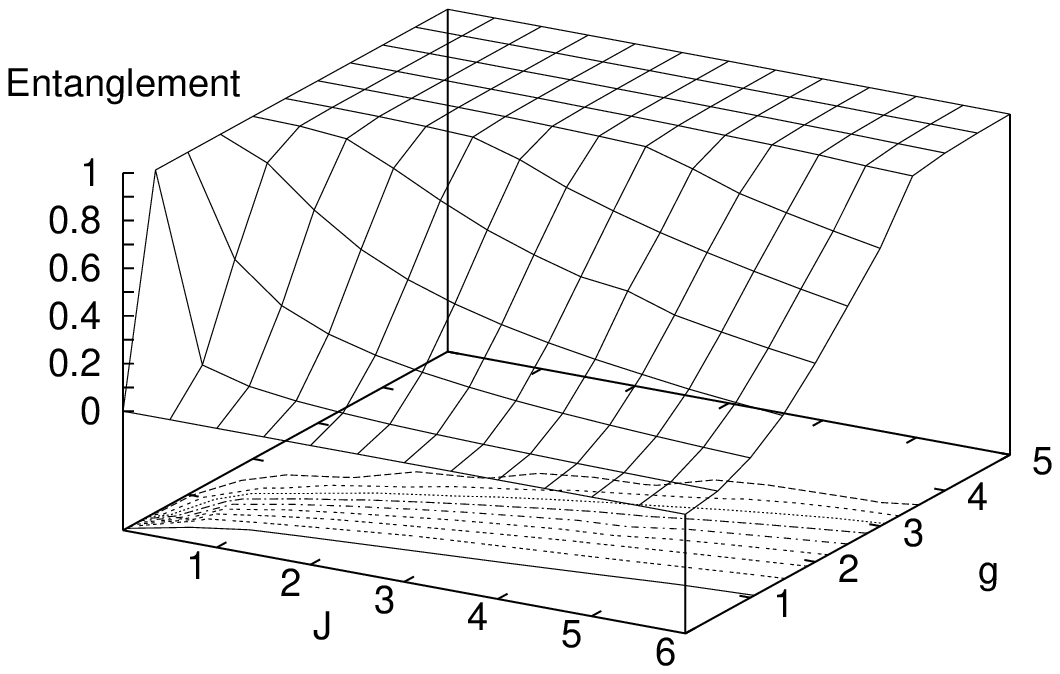}
}
\rotatebox{0}{ 
\includegraphics[scale=0.51]{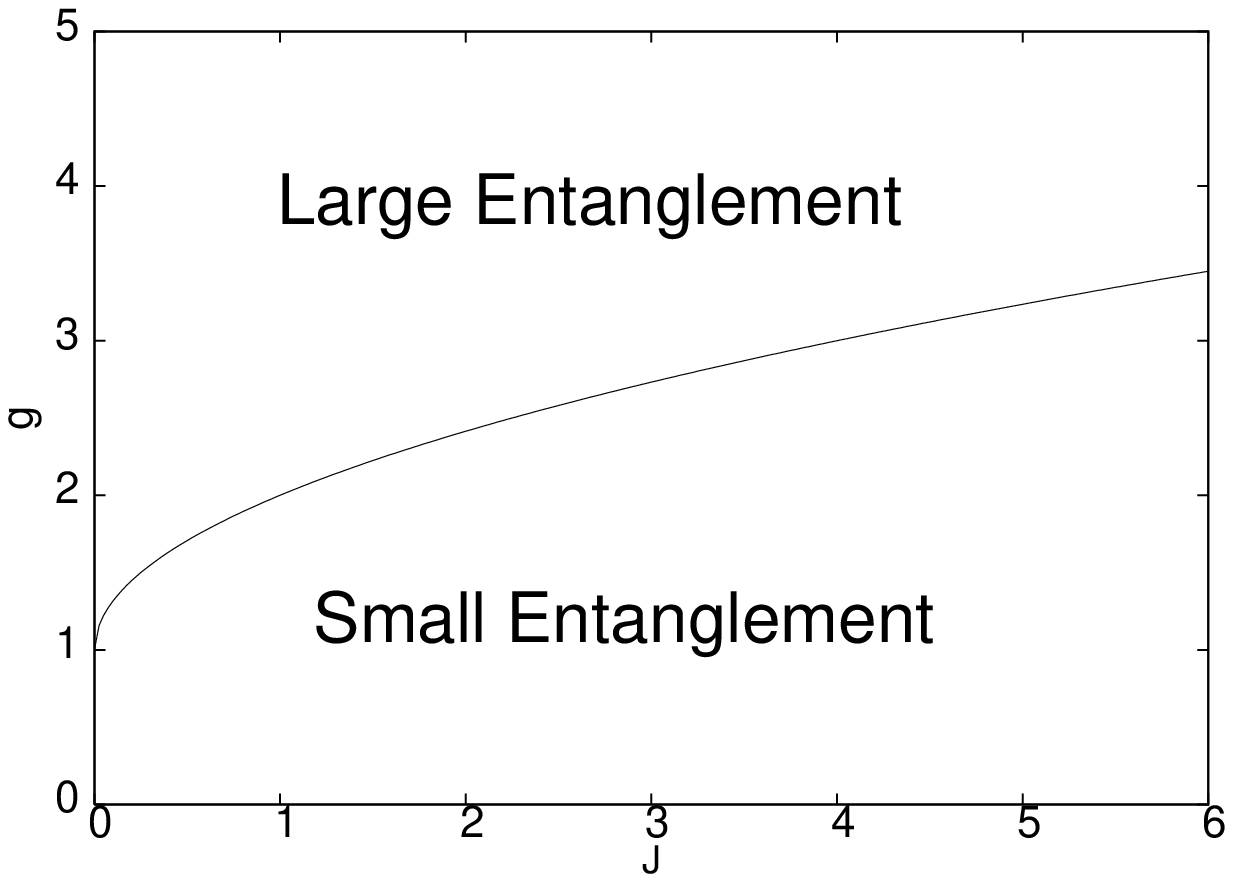}
        }
\end{center}
\caption{Hetero-entanglement between the exciton and the phonons as
measured by the linear entropy $1-{\rm Tr}_{\rm e} [( \rho_{\rm e}^{K=0})^2]$
is displayed in the upper panel for the
entire J-g phase diagram. The entanglement is calculated for the lowest energy
state with zero crystal momentum K=0. 
The solid line in the lower panel is the edge of the cliff which separates
the large and small entanglement regions. 
} \label{illus}
\end{figure}

For a finite bare exciton band width $J >  0$, the lowest 
energy state for diagonal coupling only has zero crystal momentum $K=0$.
In this lowest-energy $K=0$ state,
hetero-entanglement between the two species in
the Holstein Hamiltonian, the exciton and the phonons, as
measured by the linear entropy,
has the form
\beq
E := 1-{\rm Tr}_{\rm e} [( \rho_{\rm e}^{K=0})^2] .
\label{e14}
\enq
The function (\ref{e14}) is a linearized version of the von Neumann 
entropy $S(\rho)=-\rm{Tr} \rho \ln \rho;$ 
it shares  with this latter quantity the properties: (i) $E=0$
 $\Leftrightarrow$ $\rho=\rho^2$, i.e., $\rho$ is a pure state; (ii) 
$E_{\rm{max}}=1-1/D=E(\openone/D)$, i.e., the linear entropy
is at its maximum for the totally mixed state $\openone/D$
($D=$ is the dimension of the space). Moreover, for qubits,
the linear and von Neumann entropy are monotonic functions
of each other. 
The linear entropy (\ref{e14}) has a close relation with the
so-called 2-Renyi entropy \cite{renyi}.
These considerations show that (\ref{e14})
represent a good choice as a simple entanglement measure. 
Calculated hetero-entanglement is shown in Figure \ref{illus}
for the entire $J$ vs g phase diagram.                  
It is important to note that the qualitative behavior of the ground state
entanglement as a function of control parameters discussed above
does {\em not} depend on the specific choice of the linear entropy
as the entanglement measure; analogous calculations performed with the von
Neumann entropy give rise to the very same qualitative picture \cite{unpub}

On the $J=0$ vertical axis, the {\it exact} 
wave function of the Holstein polaron
can be written as
$\sum_n 
B_n^\dagger |0\rangle_{\rm e} 
| \Lambda_n^{\rm J=0} \rangle
,$
where 
$\langle \Lambda_m^{\rm J=0} | \Lambda_n^{\rm J=0}  \rangle  = \delta_{mn}. 
$
Therefore, the hetero-entanglement between the two species reaches its maximum
on the $J=0$ vertical axis:
$E_{\rm max} = 1- N^{-1}.$
On the horizontal axis $g=0$, on the contrary, the $K=0$ state is 
separable with respect to the two species. 

As shown in the upper panel of Figure \ref{illus}, 
exciton-phonon hetero-entanglement forms a cliff.
The edge of the cliff is found overlapping with the thin
sword-shaped area in Figure 1 (i.e., the self-trapping line), 
and can be fitted 
empirically by $J_c= (g_c - 1)^2 \hbar \omega_0$: on the cliff plateau, E approaches $1-N^{-1}$;
on the other side of the cliff edge, E decreases rapidly to zero with
increasing J or decreasing g.
In the lower panel, the solid line is the projection of the cliff edge onto the J-g
plane which separates the large and small entanglement phases. 
The usual distinction between the small and large polarons 
can thus be rephrased as follows: 
{\em the small polaron is a maximally entangled exciton-phonon entity,  while
the large polaron has much-reduced exciton-phonon hetero-entanglement.
}
The hetero-entanglement is therefore a good measure of large and small polarons and
the transition in between. 

\section{Discussions}

In this paper we have revisited 
the classic polaron problem utilizing the concept of quantum entanglement.
The exciton-phonon coupling,  described by the Holstein Hamiltonian, induces strong quantum coherences
between the two heterogeneous physical degrees of freedom. 
These coherences can be conveniently quantified by the linear entropy 
of the excitonic (or phononic) reduced density matrix. 
The latter has been explicitly computed by resorting to utilization of the
Toyozawa Ansatz.
Difficulties associated  to 
the infinite dimensionality of the 
phonon degrees of freedom can be circumvented 
by using the Toyozawa variational Ansatz to approximate
the ground-state polaronic wave function, for which an exact form
remains elusive. 
This choice of Ansatz states also allows us to carry out a large portion   
of the calculations in an analytical, conceptually transparent fashion.
The study of the entanglement behavior, as a function of the
controlling parameters
$J$ and $g$ (exciton hopping amplitude and exciton-phonon coupling respectively), 
allows a
very simple characterization of the zero-temperature phase diagram of the model. The self-trapping transition
from large to small polaron regimes can be understood in terms of a sharp increase of the exciton-phonon entanglement.

We also note that this form of hetero-entanglement is complementary to 
the superradiant behavior of the 
excitonic system. 
Superradiance (coherent spontaneous emission) is
the enhanced radiative decay compared to that of a monomer
as a result of the coherent nature of the electronic excited states 
\cite{sup,prl}.
If all transition dipoles of the $N$ monomers are parallel,
superradiance of the $K=0$ state can be calculated from
\begin{eqnarray}
S:=\sum_{mn} (\rho^{K=0}_{\rm e})_{mn} 
\end{eqnarray}
It follows that excitonic superradiance and the exciton-phonon hetero-entanglement 
are complementary.
For example, on the vertical J=0 axis in Figure~2,
the linear hetero-entanglement (\ref{e14}) is at its maximum, $1-N^{-1}$, while
the corresponding $K=0$ superradiance
reaches its minimum value 1; on the horizontal $g=0$ axis, the linear
hetero-entanglement vanishes while corresponding $K=0$ superradiance reaches its
maximum value $N$. For a given transfer integral $J$, as the exciton-phonon coupling
$g$ is reduced, the superradiance gains while the exciton-phonon
hetero-entanglement decreases.
Superradiance can be regarded as a form of the quantum coherence
between different components of 
the exciton wave function localized on different sites.
This coherence, in turn, can be viewed as a form of entanglement between
spatially distinct excitonic {\em modes} 
(for a definition of mode entanglement, see, e.g., \cite{ferm}).
Thus, the aforementioned complementarity phenomenon 
can be viewed
as a form of entanglement transfer from the inter-exciton-phonon type to 
the intra-exciton mode entanglement, and vice versa.

The role of off-diagonal exciton-phonon coupling 
as well as analyses of entanglement of polaronic states
with nonzero crystal momenta are the subject of ongoing investigations. 
The Toyozawa Ansatz is again well-suited to capture the underlying physics
in the presence of off-diagonal exciton-phonon coupling.   
We believe that the results presented in this paper 
exemplify the fact that implementations
of notions and apparatuses drawn from the emerging field 
of quantum information science are invaluable for 
gaining fresh insights for classic problems in condensed matter physics.

\section*{Acknowledgments}     

Support from the Hong Kong Research Grant Council (RGC Competitive Earmarked 
Research Grant Award HKU 7010/03P) and 
the Committee for Research and Conference Grants of the
 University of Hong Kong is gratefully acknowledged. 
P.Z. thanks Z.D.~Wang for the invitation and hospitality at the
Physics Department of University of Hong Kong.


\begin{thebibliography}{99}
\bibitem{lan}  L.D.~Landau, Phys.~Z.~Sowjetunion {\bf 3}, 644 (1933).

\bibitem{pek}  S.I.~Pekar, J.~Phys.~(Moscow) {\bf 10}, 341 (1946).

\bibitem{ent} M. Horodecki, P. Horodecki and R.
Horodecki, in {\it ``Quantum Information - Basic Concepts and
Experiments,''} Eds. G. Alber and M. Weiner,
in print (Springer, Berlin, 2000).

\bibitem{qip} For reviews, see D.P. Di Vincenzo and C. Bennet  {\sl Nature} {\bf 404}, 247 (2000);
 A. Steane, Rep. Prog. Phys. {\bf 61}, 117 (1998)

\bibitem{qpt-ent}    Tobias J. Osborne, Michael A. Nielsen, Phys. Rev. A, {\bf66}, 32110 (2002);
A. Osterloh et al,  Nature {\bf 416}, 608 (2002);  G. Vidal et al, Phys.  Rev. Lett. {\bf{90}}, 22790 (2003);
T.-C Wei et al, quant-ph/0405162 

\bibitem{comp-scheme} G. Vidal. Phys. Rev. Lett {\bf 91}, 147902 (2003), quant-ph/0310089;
M. Zwolak, G. Vidal cond-mat/0406440; F. Verstraete et al  cond-mat/0406426.  

\bibitem{munn}
R.~W.~Munn, and R.~Silbey,
{ J. Chem. Phys. \bf 83}, 1843 (1985);
{\it ibid.} {\bf83}, 1854 (1985). 

\bibitem{zhao94}
Y.~Zhao, D.~W.~Brown, and K.~Lindenberg,
J. Chem. Phys.  {\bf 100}, 2335 (1994); {\it ibid.} {\bf 106}, 2728 (1997). 

\bibitem{nago}
S.~Ishihara and N.~Nagaosa, Phys. Rev. B {\bf 69}, 144520 (2004). 

\bibitem{holstein} T. Holstein, Ann. Phys. NY {\bf 8}, 325 (1959);
{\it ibid} {\bf 8}, 343 (1959).        

\bibitem{Matsui}
A.~Matsui and K.~Mizuno, Exciton dynamics in organic molecular crystals, 5th
Intern. Conf. on Excited States in Solids, Lyon (1985).       

\bibitem{emin73}
D.~Emin, Adv.~Phys.~{\bf 22}, 57(1973).     

\bibitem{toy}
Y.~Toyozawa,  Prog. Theor. Phys. {\bf 26}, 29 (1961).

\bibitem{thesis}
Y.~Zhao, Doctoral thesis, University of California, San Diego, 1994.

\bibitem{zhaos}
Y.~Zhao, D.~W.~Brown, and K.~Lindenberg,
J.~Chem.~Phys. {\bf 107}, 3159 (1997);
{\it ibid.} {\bf 107}, 3179 (1997);
{\it ibid.} {\bf 106}, 5622 (1997); A.~Romero,
D.~W.~Brown, and K.~Lindenberg, {\it ibid.} {\bf 109}, 6540 (1998).

\bibitem{pekar}
V.M.~Buinmistrov and S.I.~Pekar, Sov. Phys. JETP {\bf 5}, 970 (1957).

\bibitem{jeck}
E.~Jeckelman and S.R.~White, Phys. Rev. B {\bf 57}, 6376 (1998).

\bibitem{trug}
J.~Bonca, S.A.~Trugman and I.~Batisti\'c, Phys. Rev. B {\bf 60}, 1633 (1999).

\bibitem{renyi}
A. Renyi, On measures of entropy and information, in:  Proc. Fourth. 
Berkeley Symp. Math. Stat. Prob. 1960, Vol. I, University of California 
Press, Berkeley, 1961, p. 547

\bibitem{unpub}
Y.~Zhao, P.~Zanardi, and G.H.~Chen, unpublished.

\bibitem{sup}
Y.~Zhao {\it et al.}, J. Phys. Chem. B {\bf 103}, 3954 (1999); T.~Meier,
Y.~Zhao, V.~Chernyak, and S.~Mukamel, J. Chem. Phys.~{\bf 107}, 3876 (1997);   
Y.~Zhao, G.H.~Chen, and L.~Yu, {\it ibid.}
{\bf 113}, No.~16, 6502 (2000).      

\bibitem{prl} M.~Lippitz {\it et al.},
Phys.~Rev.~Lett.~{\bf 92}, 103001 (2004);
S.H.~Lim {\it et al.}, {\it ibid.}~{\bf 92}, 107402 (2004);
F.~Meinardi {\it et al.}, {\it ibid.}~{\bf 91}, 247401 (2003).

\bibitem{ferm} P. Zanardi, 
Phys. Rev. A {\bf 65}, 042101 (2002).



\end{thebibliography}
\end{document}